\documentclass{article}

\usepackage{etoolbox}
\newtoggle{arxiv}
\toggletrue{arxiv} % or \togglefalse{arxiv}

\usepackage[T1]{fontenc}
\usepackage[utf8]{inputenc}

\iftoggle{arxiv}{
    \usepackage{ismir}
}{
    \usepackage[submission]{ismir} % Remove the "submission" option for camera-ready version
}

\newcommand\blfootnote[1]{%
  \begingroup
  \renewcommand\thefootnote{}\footnote{#1}%
  \addtocounter{footnote}{-1}%
  \endgroup
}

\usepackage{amsmath,cite,url}
\usepackage{graphicx}
\usepackage{color}

\usepackage{amsthm}
\usepackage{amsmath}

\usepackage{placeins}
\usepackage{url}
\usepackage{booktabs}
\usepackage{amsmath, amsfonts}
\usepackage{overpic}
\usepackage{amssymb}
\usepackage{appendix}
\usepackage{todonotes}
\usepackage{leftindex}
\usepackage{algorithm}
\usepackage{algorithmic}
\usepackage{xcolor}
\usepackage[inline]{enumitem}
\usepackage{multirow}
\usepackage{scalerel}
\usepackage{tikz}
\usetikzlibrary{3d}  % optional, for easier 3D transforms
\usepackage{tikz-3dplot}
\usepackage{algorithmic}
\usepackage{dsfont}
\usepackage{colortbl}
\usepackage{tabularx}

\usepackage{wrapfig}
\usepackage{arydshln}

\graphicspath{{figures/}}

\title{STAGE: Stemmed Accompaniment Generation through Prefix-Based Conditioning}

\multauthor
  {Giorgio Strano$^{1, \star}$ \hspace{1cm} Chiara Ballanti$^{1, \star}$ \hspace{1cm} Donato Crisostomi$^{1}$}
  {{\bf Michele Mancusi$^{1}$ \hspace{1cm} Luca Cosmo$^{2}$ \hspace{1cm} Emanuele Rodolà$^{1}$}\\
  $^{1}$Sapienza University of Rome, $^{2}$Ca’ Foscari University of Venice\\
  {\tt\small strano@di.uniroma1.it}
  }

\def\authorname{F. Author, S. Author, and T. Author}

\definecolor{myblue}{rgb}{0.21,0.49,0.74}

\iftoggle{arxiv}{
    \usepackage[bookmarks=false,pdfsubject={\pdfsubject},hidelinks, colorlinks,allcolors=myblue]{hyperref}
}{
    \usepackage[bookmarks=false,pdfauthor={\authorname},pdfsubject={\pdfsubject},hidelinks]{hyperref}
}

\newcommand{\musicgen}{\texttt{MusicGen}}
\newcommand{\musiclm}{\texttt{MusicLM}}

\newcommand{\musicongen}{\texttt{MusicConGen}}
\newcommand{\musicgenstem}{\texttt{MusicGen-Stem}}
\newcommand{\instructmusicgen}{\texttt{Instruct-MusicGen}}
\newcommand{\encodec}{\texttt{EnCodec}}
\newcommand{\jukebox}{\texttt{Jukebox}}

\newcommand{\msdm}{\texttt{MSDM}}
\newcommand{\gmsdi}{\texttt{GMSDI}}
\newcommand{\sacontrolnet}{\texttt{SA-ControlNet}}
\newcommand{\stemgen}{\texttt{StemGen}}
\newcommand{\diffariff}{\texttt{Diff-A-Riff}}

\makeatletter
\def\adl@drawiv#1#2#3{%
        \hskip.5\tabcolsep
        \xleaders#3{#2.5\@tempdimb #1{1}#2.5\@tempdimb}%
                #2\z@ plus1fil minus1fil\relax
        \hskip.5\tabcolsep}
\newcommand{\cdashlinelr}[1]{%
  \noalign{\vskip\aboverulesep
           \global\let\@dashdrawstore\adl@draw
           \global\let\adl@draw\adl@drawiv}
  \cdashline{#1}
  \noalign{\global\let\adl@draw\@dashdrawstore
           \vskip\belowrulesep}}
\makeatother

\newcommand{\vertical}[1]{\rotatebox[origin=c]{90}{#1}}

\newcommand{\stage}{\texttt{STAGE}}
\begin{document}

\maketitle

\begin{abstract}
    Recent advances in generative models have made it possible to create high-quality, coherent music, with some systems delivering production-level output. \blfootnote{$\star$ denotes equal contribution.}
    % problem
    Yet, most existing models focus solely on generating music from scratch, limiting their usefulness for musicians who want to integrate such models into a human, iterative composition workflow. 
    % solution (with approach details)
    In this paper we introduce \stage{}, our \emph{STemmed Accompaniment GEneration} model, fine-tuned from the state-of-the-art \musicgen{} to generate single-stem instrumental accompaniments conditioned on a given mixture. Inspired by instruction-tuning methods for language models, we extend the transformer's embedding matrix with a \emph{context token}, enabling the model to attend to a musical context through prefix-based conditioning.
    % results
    Compared to the baselines, \stage{} yields accompaniments that exhibit stronger coherence with the input mixture, higher audio quality, and closer alignment with textual prompts. 
    % solution (2)
    Moreover, by conditioning on a metronome-like track, our framework naturally supports tempo-constrained generation, achieving state-of-the-art alignment with the target rhythmic structure--all without requiring any additional tempo-specific module.
    % conclusions
    As a result, \stage{} offers a practical, versatile tool for interactive music creation that can be readily adopted by musicians in real-world workflows.

    \begin{center}
        \footnotesize
        \hspace{0.2cm}\raisebox{-0.2\height}{\includegraphics[width=1em,height=1em]{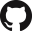}}\hfill \href{https://github.com/github.com/giorgioskij/stage}{\texttt{github.com/giorgioskij/stage}}\hspace{0.2cm}

         \hspace{0.2cm}\raisebox{-0.2\height}{\includegraphics[width=1em,height=1em]{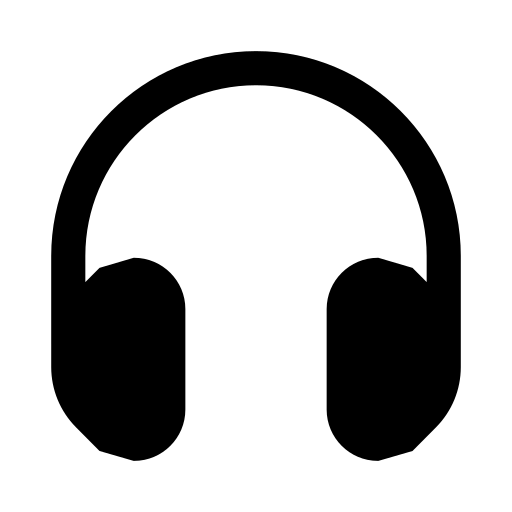}}\hfill \href{https://giorgioskij.github.io/stage-demo/}{\texttt{giorgioskij.github.io/stage-demo}\hspace{0.2cm}}
    \end{center}
\end{abstract}

\section{Introduction}
\label{sec:introduction}
Generative AI has recently transformed music composition, with large-scale models now able to produce long-form, high-quality, and stylistically consistent music. Models such as \musicgen{} \cite{musicgen}, \texttt{MusicLM} \cite{agostinelli2023musiclm}, and \texttt{JukeBox} \cite{jukebox} have demonstrated that transformers trained on tokenized audio representations can generate music that rivals human compositions in coherence and production quality.

However, most of these models focus on generating music from scratch, even when they allow for conditional generation using melodies \cite{musicgen}, chords \cite{lanmusicongen, gao2024end, 9376975, li2023melodydiffusion, jung2024musicgen}, or text prompts. This limits their applicability in a natural music composition workflow, which is often structured in an iterative, layered fashion, gradually building compositions by adding or refining parts over time.
To support this workflow, we focus on a human-centered, intuitive generation task: \textit{adding a single new stem to an existing multi-stem mixture}, while also allowing for precise control over the tempo of the generated output.

\begin{figure}
\label{fig:teaser}
    \centering
    \includegraphics[width=\linewidth]{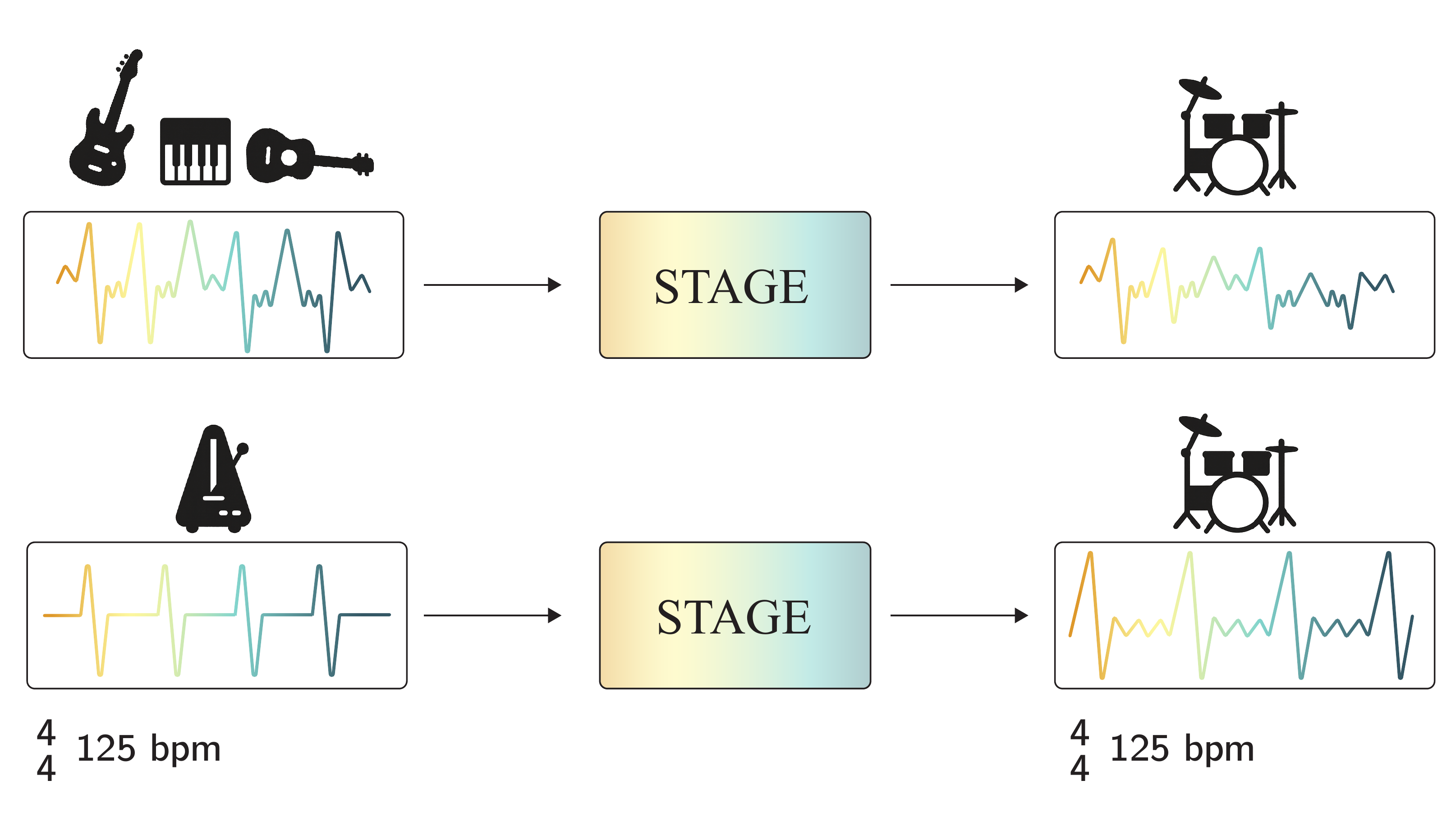}
    \caption{Outline of our proposed model. (top) \texttt{STAGE} takes a musical context as input and generates a single-stem accompaniment. (bottom) \texttt{STAGE} takes a metronome-like track and generates a stem that follows the desired rhythmic structure.}
    \label{fig:enter-label}
\end{figure}

In this paper, we introduce \stage{}, a single-instrument accompaniment generation model that can be conditioned on any audio context, be it a mixture or a simple click track, to generate a coherent and rhythmically aligned accompaniment (see Figure \ref{fig:teaser}).
We use a simple yet effective approach to fine-tune \musicgen{} \cite{musicgen} for stemmed accompaniment generation. Our method does not require retraining any additional context-encoders and relies on minimal data for fine-tuning. We leverage prefix-based conditioning, where the context is prepended to the model’s input sequence, effectively serving as a prompt for the generation of the target stem. This approach draws inspiration from instruction tuning \cite{wei2022finetuned, zhang2024instructiontuninglargelanguage} in language models, where prepending a task-specific instruction enables a pretrained model to specialize in new tasks with minimal modifications. In our case, we treat musical contexts as the ``question'' and the desired accompaniment as the ``answer''. This enables the model to learn a token-to-token correspondence between context and continuation, specializing it for accompaniment tasks.

We evaluate our model on musical coherence using the COCOLA score \cite{ciranni2025cocola}, showing clear improvements over existing baselines, while maintaining high audio quality as measured by FAD \cite{kilgour2019frechet} and KAD \cite{chung2025kad}. Furthermore, we show that our model supports tempo-constrained generation by simply conditioning on a metronome-like click track—without the need for tempo-specific modules or architectures.

Our contributions are three-fold:
\begin{itemize}
    \item A prefix-based fine-tuning method for stemmed accompaniment generation.
    \item A lightweight, flexible approach to tempo conditioning through audio-based inputs.
    \item Extensive evaluation across musical coherence, rhythmic alignment, and audio fidelity, along with open-source code and model checkpoints.
\end{itemize}

\section{Related Work}
\label{sec:related_work}
Recent advances in generative modeling treat music as a language of discrete tokens, enabling long-context audio synthesis via transformer architectures. \texttt{Jukebox} \cite{jukebox} was an early example: it converted raw audio into a hierarchy of residual VQ-VAE tokens and used progressively deeper transformers to generate coherent extended sequences of music.
Subsequent improvements in \emph{neural audio codecs}, such as \texttt{SoundStream} \cite{soundstream2021zeghidour} and \texttt{EnCodec} \cite{défossez2022highfidelityneuralaudio}, inspired new designs. \texttt{MusicLM}\cite{agostinelli2023musiclm} introduced a hierarchical two-stage approach that models separate streams of ``semantic'' and ``acoustic'' tokens, while \musicgen{} \cite{musicgen} showed that a single-stage transformer over \texttt{EnCodec} tokens can achieve state-of-the-art text-to-music quality. The simpler architecture of \texttt{MusicGen}, combined with its robust audio fidelity generations, make it a natural foundation for specialized tasks such as single stem accompaniment, which we explore in this paper.

\subsection{Conditional generation}
Although most music LMs focus on text prompting, some approaches provide more direct musical guidance or editing capabilities. Melody-conditioned models, such as the melody variant in~\cite{musicgen}, align the generation to a guiding pitch contour but typically produce an entire mix rather than an isolated stem. \musicongen{}~\cite{lanmusicongen} further extends \musicgen{} by adding conditioning over chords and beat information, allowing explicit control of harmonic and rhythmic structures. Multiple other systems have been presented, especially using diffusion models, to condition music generation on a series of chords \cite{ gao2024end, 9376975, li2023melodydiffusion, jung2024musicgen}, on stylistic references \cite{rouard2024audio}, or even on a combination of text, style, and a reference drums track \cite{tal2024joint}.

\subsection{Music editing}
Recent approaches to audio-domain editing include autoregressive models like \texttt{Instruct-MusicGen}\cite{zhang2024instruct}, as well as diffusion-based systems such as \texttt{MSDM}\cite{mariani2024multi} and \texttt{GMSDI}~\cite{postolache2024generalized}. While \texttt{Instruct-MusicGen} struggles to achieve high-quality outputs, diffusion-based models, despite their flexibility, require significantly more computational resources and do not consistently support clean, single-stem generation.

\subsection{Accompaniment generation}
Similarly to \msdm{} and \gmsdi{}, other diffusion-based systems aim to generate or edit partial arrangements but are closed-source or limited in scope. For instance, \texttt{Diff-A-Riff}~\cite{nistaldiff} uses a multi-step diffusion process to refine an existing track with new musical elements; however, it is not openly released. \texttt{SA-ControlNet}\footnote{\href{https://github.com/EmilianPostolache/stable-audio-controlnet}{github.com/EmilianPostolache/stable-audio-controlnet}} uses a fine-tuning of \texttt{Stable Audio Open}~\cite{evans2025stable} with an added \texttt{ControlNet} module\cite{Zhang_2023_ICCV}, also providing a form of stemmed accompaniment generation. \texttt{SingSong}~\cite{singsong} tackles vocal-to-instrumental accompaniment, taking a vocal track as input and generating a band-like backing. This approach is highly effective for vocals but remains highly specialized.
More directly aligned with our objectives, \texttt{StemGen}\cite{parker2024stemgen} enables single-stem accompaniment generation via a non-autoregressive transformer. In parallel to our work, Meta AI introduced \texttt{MusicGen-Stem}\cite{musicgen-stem}, which supports a range of editing tasks, including mixture-conditioned accompaniment generation. However, neither model has released public code or checkpoints at the time of writing, making direct comparison difficult. Additionally, both approaches involve training dedicated transformer models from scratch, in contrast to our lightweight fine-tuning strategy.

% \textbf{MusicGen-Stem} \cite{musicgen-stem} can perform various editing tasks, including accompaniment generation by conditioning on a generic mixture, overlapping with the goal of this research. This method was developed concurrently with our work and has only recently been accepted to ICASSP2025. Its model is currently not open-source, making direct, objective re-evaluation or comparison impossible. Moreover, it requires a full pre-training of the language model from scratch. A similar scenario arises with \textbf{StemGen} \cite{9}, which offers single-stem accompaniment but is also non–open-source and provides only a few sample outputs on its website, limiting objective assessment.

% Our method, \textbf{STAGE}, is precisely aimed at this broader requirement. It offers a simple yet powerful mechanism to prepend any audio mixture, and optionally a beat track, to a large music LM, thereby enabling single-stem generation aligned to both harmonic context and precise rhythmic structure. Crucially, STAGE does not rely on symbolic analysis (e.g., chord detection) or dedicated external modules, allowing it to generalize across varied musical genres, tempos, and production workflows.

\section{Background}
\label{sec:background}
\texttt{MusicGen} \cite{musicgen} is a single-stage music generation model that operates over discrete audio tokens produced by an encoder–decoder neural codec. Specifically, the authors use \texttt{EnCodec} \cite{défossez2022highfidelityneuralaudio}, which converts raw audio into several parallel streams of quantized tokens (known as codebooks). Whereas some prior works (e.g., \jukebox{} \cite{jukebox}, \musiclm{} \cite{agostinelli2023musiclm}) rely on multi-stage or hierarchical architectures that process one set of tokens to then upsample another, \musicgen{} proposes a simpler yet effective single-stage transformer language model that directly learns to generate all of these quantized tokens at once.

\subsection{Architecture overview}
The core of \musicgen{} is a GPT-like transformer decoder that is trained autoregressively over sequences of discrete audio tokens. The tokens come from a residual vector quantization scheme, where the raw waveform is first encoded into a low-frame-rate continuous representation, and then each frame is quantized by multiple ``stacked'' codebooks. Codebooks are organized hierarchically, and each codebook $k_i$ contains incremental residual information w.r.t. the previous codebooks $k_j, i>j$. The number of codebooks (set to 4 in \musicgen{}) determines how many parallel tokens must be modeled at each time step.

\musicgen{} is released in four versions, \texttt{Small} ($\sim$400M params), \texttt{Medium} ($\sim$1.5B params), \texttt{Large} ($\sim$3B) and \texttt{Melody} ($\sim$1.5B). The latter uses both text and a short audio clip (from which melody is extracted) as conditioning, with both sources prepended to the input. The other three rely only on text, provided via cross-attention. In this work, we use only the pre-trained \texttt{MusicGen Small} checkpoint.

\subsection{Codebook Interleaving Patterns}
One straightforward approach to convert the four parallel streams of tokens generated by \texttt{EnCodec} into a single stream is to flatten all codebooks, but this substantially lengthens the sequence. Conversely, predicting them in parallel underestimates cross-codebook dependencies. A practical compromise, used in MusicGen and shown in Figure \ref{fig:codebook}, is the ``delay'' strategy which shifts tokens from the $i$th codebook by $i-1$ steps. This preserves inter-codebook context while requiring only one autoregressive step per frame. As a result, the delay pattern yields more efficient modeling than parallel prediction with minimal computational overhead.

\section{Method}
\label{sec:method}
In this section, we present the key components of \stage{}. In particular, we discuss how it extends \musicgen{}'s architecture and training procedure to generate single stems from an input mixture or a metronome-like beat track. 
We refer to both these audio inputs as the ``context''. During inference, the user can pass either of those audio cues or even combine them into a single waveform, allowing to perform accompaniment generation with an even tighter control over the target's rhythmic structure.

% We introduce \textbf{\stage{}} as a fine-tuned extension of MusicGen \cite{musicgen}, leveraging a minimal prefix-based modification to handle single-stem generation from an existing musical context. We adopt MusicGen’s \textbf{delay codebook interleaving pattern}, which diagonally offsets the tokens for each codebook stream by one timestep. Below, we describe how we incorporate an additional \textbf{context token} into this interleaved sequence and outline our training procedure, including tempo conditioning and data augmentation.

\subsection{Overview}
\label{subsec:overview}
Having chosen a target instrument $I$, \stage{} is trained to produce a stem $S$ of that instrument, taking as input:
\begin{itemize}
    \item An audio context, which can contain a generic audio mixture $M$ (without our target instrument $I$), \textbf{or} a beat track $B$ (a metronome-like pulse sequence);
    \item an optional text prompt $T$, describing the desired style and mood of the generation.
\end{itemize}

The model aims to generate audio that matches the mixture’s key, style, harmony, and rhythm, enabling it to serve as a musical accompaniment. If no initial mixture is available, \stage{} can instead take a metronome track ($B$) (a simple beat marking the tempo) as input, allowing it to generate the first stem of a composition.
In practice, we find that overlaying the mixture $M$ and beat track $B$ into a single audio file at inference time allows \stage{} to condition on both, preserving coherence with the mixture while achieving tighter rhythmic alignment with the beats (see \ref{subsec:combining} for results).

% We train separate \stage{} models for different target instruments: drums, bass, guitar, and piano/keyboards. Due to data scarcity in guitar and piano tracks in the dataset we used for training, we only present final results for \emph{drums} and \emph{bass}.
We train separate \stage{} models for each target instrument, specifically \emph{drums} and \emph{bass}. For the remaining, less-represented instruments, the inherent intra-instrument variability would require significantly more data than what is available in MoisesDB.

\subsection{Context token for prefix-based conditioning}
\label{subsec:contexttoken}
\begin{figure*}[t]

    \centering
    \includegraphics[width=\linewidth]{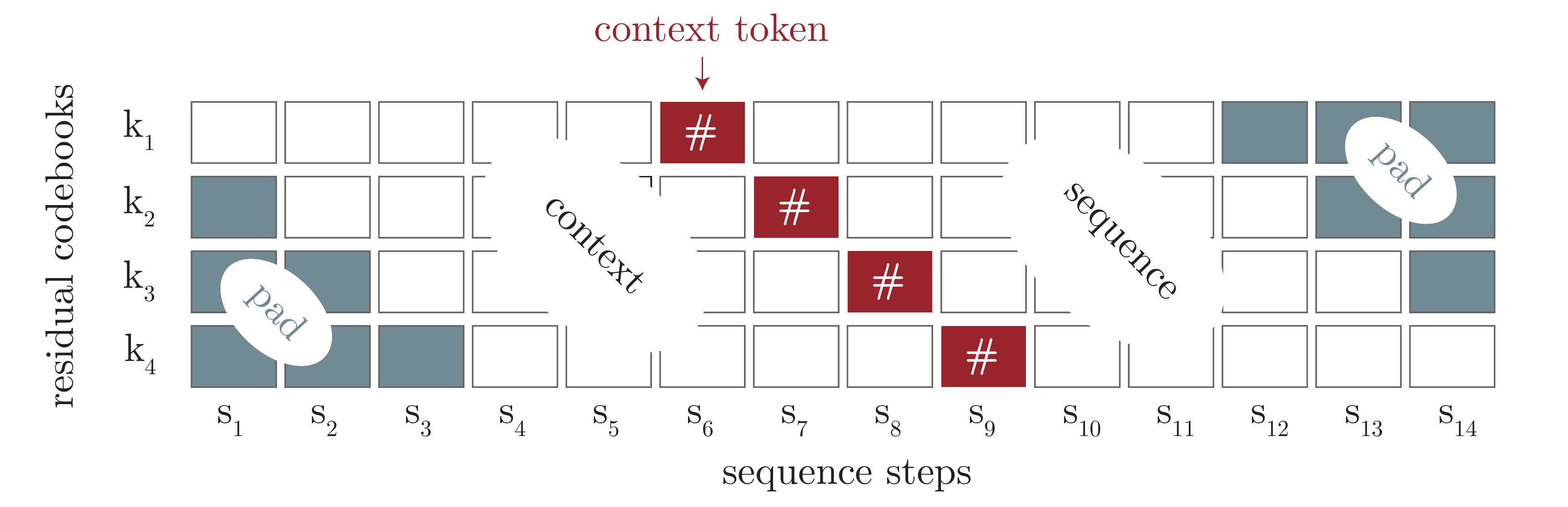}
    \caption{Illustration of the \textit{delay} pattern used by MusicGen, and how the context token is placed to separate the audio context from the input sequence of the transformer.}
    \label{fig:codebook}
\end{figure*}

Our starting point is \texttt{MusicGen-Small}, a lightweight variant of \musicgen{}. To enable conditioning, we add a single \textit{context token} to the transformer's embedding matrix, allowing extra audio tokens to be prepended to the input sequence (Figure \ref{fig:codebook}). These tokens come from either the mixture $M$ or the beat track $B$, encoded via \encodec{} \cite{défossez2022highfidelityneuralaudio}. This forms a prefix-based conditioning setup: once the model `'sees'' the context tokens, it autoregressively generates the new stem.

\subsection{Fine-tuning procedure}
\label{subsec:training}
We train on the open-source, multi-stem dataset {MoisesDB}\cite{moisesdb}, which contains 240 stem-separated songs. Given a target instrument $I$ to be generated, we create input data for \stage{} using the following strategies:
\begin{itemize}
    \item \textbf{Form the context.} For each track we either (a) mix a random subset of stems (excluding instrument $I$) to create $M$, or (b) replace the mixture with a metronome track $B$ at the known tempo of $M$. We use each of the two strategies with equal probability.  
    %\item \textbf{Set the target.} The withheld single stem (bass, drums, \emph{etc.}) is the generation target $S$.
    \item \textbf{Context length.} We randomize the context length in the range of 5 to 10 seconds. Hence, the model can learn to generate samples longer than the actual context window.
    \item \textbf{Data augmentation.} We apply speed transposition (in the range [0.8, 1.2]) and pitch transposition (in the range [-4, +4 semitones]) with probability 0.5 to both context and target. 
\end{itemize}
% The above procedure yields <context, target stem> pairs, which we feed into a minimal fine-tuning loop. For the first 200 steps, we train only the new embedding for the \textit{context token} at a learning rate of 1e-4, while all other parameters remain frozen.
% % With cross-entropy on $S$'s tokens, we typically converge in $\approx 1{,}000$ steps. We freeze most of the \texttt{MusicGen-Small} weights initially, then gradually unfreeze. 
% Subsequently, we unfreeze the rest of the model, ramping its learning rate from 0 to 1e-5 while lowering the embedding’s rate from 1e-4 to 1e-5. Training converges after approximately 1,000 steps, with batches of 8 10-second samples. The complete fine-tuning procedure amounts to less than a day on a single NVIDIA RTX 3090 GPU.
The above procedure produces <context, target stem> pairs, which we use for fine-tuning. To allow the newly introduced \textit{context token} to adapt to the pretrained model, we first train only its embedding for 200 steps at a learning rate of 1e-4, keeping all other weights frozen. This warm-up phase helps the token learn a meaningful interface with the rest of the model. We then unfreeze the remaining weights and gradually ramp up their learning rate from 0 to 1e-5, while annealing the context token’s rate from 1e-4 to 1e-5. Training converges in about 1,000 steps using batches of eight 10-second samples, finishing in under a day on a single NVIDIA RTX 3090 GPU.

\subsection{Inference}
At inference time, we perform the following steps:
\begin{itemize}
    \item Tokenize the context via \encodec{}.
    \item Pass the encoded tokens to \stage{}, followed by the {context token}.
    \item Autoregressively decode the new stem’s tokens.
    \item Reconstruct the generated stem via \encodec{}’s decoder.
\end{itemize}
This lightweight prefix-based framework is easy to integrate into a musician’s workflow by simply providing an audio snippet of an existing track (or a metronome pulse in its place) and letting \stage{} generate a single, coherent accompaniment stem.

\section{Experiments and Results}
\label{sec:experiments}
We now present the experimental setup and evaluation metrics used to assess the performance of \stage{}. We focus our evaluation on two separate tasks: 
\begin{itemize}
    \item \textit{Beat Alignment}, given a beat as conditioning;
    \item \textit{Accompaniment Generation}, given a generic mixture of stems as conditioning.
\end{itemize}
For each of these tasks, we measure against state-of-the-art open-source models on comparable tasks. 

\subsection{Beat alignment}

To measure how precisely \stage{} follows the given beat, we provide the model with a raw pulse track spanning the full duration of the sample to be generated. We then evaluate the {F1 score} using the \texttt{mireval}\footnote{\href{https://github.com/mir-evaluation/mir_eval}{github.com/mir-evaluation/mir\_eval}} library, matching the detected beats in the output audio to the reference beat track supplied as input. For beat detection, we employ Beat-This\cite{foscarinbeat}, a state-of-the-art algorithm for beat tracking.

We benchmark \stage{} against \musicongen{}~\cite{lanmusicongen}, a fine-tuned \musicgen{} variant that can be conditioned on tempo, and optionally on chord sequences. 
% To match their evaluation procedure, we supply only a beat track to our model, then measure F1 alignment scores between the supplied beats and the generation.

Following the setup of \musicongen{}, we evaluate our generations by conditioning on beat tracks extracted from MusDB~\cite{musdb18} mixtures using the Beat-This algorithm. However, we find that this process introduces noise, as the beat tracker is imperfect and often detects irregular or inconsistent beat patterns across samples.
This is not in line with a realistic scenario, in which a musician supplies the model with a perfectly regular beat grid to follow. For a more realistic setting, we also test our model on a uniform distribution of BPMs in the interval [100, 180]. 

Table \ref{tab:main-results-rhythm} shows that the model trained on drums significantly outperforms both \musicongen{} and our bass-trained variant. This supports the intuitive notion that training on drums gives the model a stronger sense of rhythm, leading to better alignment. Additionally, beat extraction is less accurate on bass-only tracks (like those generated by \stage{}\texttt{-bass}) which can contribute to higher measured alignment error.

\renewcommand{\arraystretch}{1.2}
\begin{table}[h!]
    \centering
    \resizebox{\columnwidth}{!}{
    \begin{tabular}{llccc}
        \toprule
        \textbf{Model} & \textbf{Dataset} & F1 $\uparrow$ & FAD-VGGish $\downarrow$ & FAD-Clap $\downarrow$ \\
        \midrule
        \multirow{2}{*}{\centering STAGE-drums}
          & MusDB        & 66.88  & 1.40  & 0.23 \\
          & Uniform BPM  & 71.57  & 2.05  & 0.24 \\
        \midrule
       \multirow{2}{*}{\centering STAGE-bass}
          & MusDB        & 40.93  & 5.59  & 0.39 \\
          & Uniform BPM  & 45.17  & 4.26  & 0.52 \\
        \midrule
        MusiConGen-Tempo
          & MusDB        & 61.37  & 1.95 & -- \\
        \bottomrule
    \end{tabular}
    }
    \caption{Comparison of audio quality (FAD) and rhythmic alignment (F1) of \stage{}\texttt{-drums} vs.\ \musicongen{} (with Rhythm-only conditioning). The rhythm conditioning was extracted with Beat-This from the MusDB dataset. For our model, we also test on 160 samples from a uniform distribution of BPMs in the range [100, 180]. Following \cite{lanmusicongen}, FAD is computed with MusDB as reference.}
    \label{tab:main-results-rhythm}
\end{table}

\renewcommand{\arraystretch}{1.2}
\begin{table*}[h!]
    \resizebox{1\textwidth}{!}{
        % dataset approach CoCoLa FAD-vggish FAD-clap KAD-vggish KAD-clap KL SIMM
        % c c c cccccc
        \begin{tabular}{cc c cccccc}
            %
            %%%%%%%%%%%%%%%%%%%%%%%%%%%%%%%%%%%%%%%%%%%%%%%%%%%%%% HEADER %%%%%%%%%%%%%%%%%%%%%%%%%%%%%%%%%%%%%%%%
            %
            \toprule
                                                           &                          & COCOLA     & FAD-VGGish   & FAD-Clap     & KAD-VGGish   & KAD-Clap       & Rhythmic Alignment (F1)       \\
            \multirow{-2}{*}{Target stem}                  & \multirow{-2}{*}{Model} & $\uparrow$ & $\downarrow$ & $\downarrow$ & $\downarrow$ & $\downarrow$ & $\uparrow$ \\

            \cmidrule(lr){1-1} \cmidrule(lr){2-2} \cmidrule(lr){3-9}
                                                           % & \musicgen{}              & --         & --           & --           & --           & --             & --         \\
                                                           & \texttt{STAGE}           & \textbf{61.02}& \textbf{1.05} & \textbf{0.17}         & \textbf{3.00}    & \textbf{9.86} & \textbf{52.63}       \\
                                                           & \instructmusicgen{}      & 48.24      & 16.44        & 1.34         & 55.25        & 80.79          & 0.20      \\
                                                           & \texttt{GMSDI}           & 45.25      & 15.57         & 1.22         & 57.39           & 54.75             & 25.98         \\
            \multirow{-4}{*}{\small \vertical{Drums}}      & \texttt{SA-ControlNet}     & 57.46      & 2.75         & 0.39         & 9.28           & 11.75             & 38.70         \\
            \midrule
                                                           % & \musicgen{}              & --         & --           & --           & --           & --             & --         \\
                                                           & \texttt{STAGE}           & \textbf{60.20} & 2.97         & 0.37         & 13.45           & \textbf{15.40}    & 40.94         \\
                                                           & \instructmusicgen{}      & 53.69         & 12.85           & 1.28    & 47.41           & 73.92             & 0.19         \\
                                                           & \texttt{GMSDI}           & 43.29      & 14.36         & 1.28         & 44.31           & 49.55              & 24.34         \\
            \multirow{-4}{*}{\small \vertical{Bass}}       & \texttt{SA-ControlNet}     & 59.63      & \textbf{2.22}  & \textbf{0.31}  & \textbf{6.69}  & 15.98           & \textbf{47.17 }          &         \\
            \bottomrule
        \end{tabular}
    }
    \caption{Performance comparison on the accompaniment generation task using the MoisesDB test set, with \emph{bass} and \emph{drums} as target stems.}
    \label{tab:main-results-accompaniment}
\end{table*}

\subsection {Accompaniment coherence}
To assess how well our model adds a new instrument stem to an existing mixture, we evaluate on the test set from the MoisesDB dataset \cite{moisesdb}. For each track, we remove the target stem (drums or bass) and then ask the model to regenerate that instrument while keeping the remaining parts unchanged. We measure several metrics that target both objective audio quality and semantic coherence with the context:
\begin{itemize}
    \item COCOLA score~\cite{ciranni2025cocola} captures harmonic and percussive coherence between the context and the newly generated stem;
    \item FAD-vggish and FAD-clap \cite{kilgour2019frechet} assess perceptual quality by comparing the distribution of embeddings (extracted using VGGish~\cite{hershey2017cnn} or CLAP~\cite{elizalde2023clap}) between generated and reference audio. Since FAD measures distributional distance, we use stems from MoisesDB matching the target instrument to define the reference. 
    \item KAD-vggish and KAD-clap~\cite{chung2025kad}. A newly released metric that computes the distance between distributions of embeddings in a higher-order abstract space, using the kernel trick. It has similar properties to the commonly used FAD metric.
    \item Rhythmic Alignment (F1). We also compute the F1-score between beats extracted from the context, and beats extracted from the generated stem, to asses the rhythmic alignment and coherence between context and accompaniment.
\end{itemize}

The scores for \gmsdi{} and \sacontrolnet{} are computed on samples provided by the authors, while Instruct-MusicGen was run locally with the public inference code\footnote{\href{https://github.com/ldzhangyx/instruct-MusicGen}{github.com/ldzhangyx/instruct-MusicGen}}. We were not able to compare with \stemgen{}~\cite{parker2024stemgen}, \diffariff{}~\cite{nistaldiff}, and \musicgenstem{}~\cite{musicgen-stem} since their code is not publicly available.

% Table \ref{tab:main-results-accompaniment} shows how our proposed model is able to outperform all baselines in semantic coherence with the context, audio quality, and rhythmic alignment when generating drums, and performs comparably to \sacontrolnet{} on bass.  In general, we observe that \stage{} exhibits slightly weaker performance on all bass metrics than drums. We impute this to:
% \begin{enumerate}[label=\alph*)]
%     \item A higher variety in the distribution of bass tracks in the MoisesDB dataset. Indeed, bass tracks from the dataset include both electric and synth bass, which sound very different from each other; 
%     \item A higher intrinsic difficulty of the generation of bass tracks. Differently from the drums, the bass has to take into account both rhythmic and harmonic information from the context, creating a harder distribution to model.
% \end{enumerate}

As shown in Table~\ref{tab:main-results-accompaniment}, our proposed model outperforms all baselines in semantic coherence, audio quality, and rhythmic alignment when generating \emph{drums}, and performs on par with \sacontrolnet{} for \emph{bass}. Overall, we observe that \stage{} performs slightly worse on bass across all metrics. We attribute this to two main factors: 
\begin{enumerate}[label=\alph*)] 
    \item Greater variability in the distribution of bass tracks within the MoisesDB dataset, which includes both electric and synth bass with distinct timbral characteristics. 
    \item The inherently more complex nature of bass generation, which--unlike drums--requires modeling both rhythmic and harmonic information from the context, resulting in a more challenging prediction task. 
\end{enumerate}

\subsection{Ablation study on rhythm conditioning}
To evaluate the impact of training on <metronome, target> pairs alongside <mixture, target> pairs, we conduct an ablation study (Table \ref{tab:main-results-ablation}). We fine-tune a version of \stage{}\texttt{-drums} without metronome tracks, exposing it only to <mixture, target> pairs. As expected, the model fails to generate tempo-aligned outputs when conditioned on a metronome at inference. More notably, its performance \emph{also degrades on accompaniment generation} (the very task it was trained for) highlighting the broader benefit of including metronome conditioning at training.

\renewcommand{\arraystretch}{1.2}
\begin{table*}[h!]
\resizebox{1\textwidth}{!}{
        % dataset approach CoCoLa FAD-vggish FAD-clap KAD-vggish KAD-clap KL SIMM
        % c c c cccccc
        \begin{tabular}{cc c cccccc}
            %
            %%%%%%%%%%%%%%%%%%%%%%%%%%%%%%%%%%%%%%%%%%%%%%%%%%%%%% HEADER %%%%%%%%%%%%%%%%%%%%%%%%%%%%%%%%%%%%%%%%
            %
            \toprule
                                                           &                          & COCOLA     & FAD-VGGish   & FAD-Clap     & KAD-VGGish   & KAD-Clap       & Rhythmic Alignment (F1)       \\
            \multirow{-2}{*}{Target stem}                  & \multirow{-2}{*}{Model} & $\uparrow$ & $\downarrow$ & $\downarrow$ & $\downarrow$ & $\downarrow$ & $\uparrow$ \\

            \cmidrule(lr){1-1} \cmidrule(lr){2-2} \cmidrule(lr){3-9}
                                                           % & \musicgen{}              & --         & --           & --           & --           & --             & --         \\
                                                           & \texttt{STAGE}           & \textbf{61.02}& \textbf{1.05} & 0.17         & \textbf{3.00}    & 9.86 & \textbf{52.63}       \\   
            \multirow{-2}{*}{\small Drums}      & \texttt{STAGE-abl.}     & 59.28          & 1.16            & 0.17         & 4.37           & \textbf{9.59}             & 48.05         \\
            \bottomrule
        \end{tabular}
        }
    \caption{Ablation results: the two models are exactly the same, but \texttt{STAGE-abl} never sees <metronome, target> pairs during fine-tuning. Models are tested on accompaniment generation, generating drums given a mixture of stems extracted from the test set of MoisesDB. We observe that the model which observed conditioning with metronome tracks is able to better align to the rhythmic structure of the mixture, even on the normal accompaniment generation task.}
    \label{tab:main-results-ablation}
\end{table*}

% The clear difference (see Table \ref{tab:main-results-ablation}) in both COCOLA score (which evaluates rhythmic, as well as harmonic coherence) and the F1-score measured on rhythmic alignment means that including <metronome, target> pairs during fine-tuning not only makes the model able to perform tempo-controlled generation, but it also improves the model's ability to understand and reproduce rhythmic structures in the general accompaniment generation task. 
The clear gap in both COCOLA scores (which capture rhythmic and harmonic coherence) and F1 scores for rhythmic alignment (Table~\ref{tab:main-results-ablation}) indicates that including <metronome, target> pairs during fine-tuning not only enables tempo-constrained generation, but also enhances the model's overall ability to perceive and reproduce rhythmic structure in standard accompaniment generation.

\subsection{Combining mixture and metronome for improved alignment}
\label{subsec:combining}

\begin{figure}
    \centering
    \hspace{-1.5cm}\includegraphics[width=.9\linewidth]{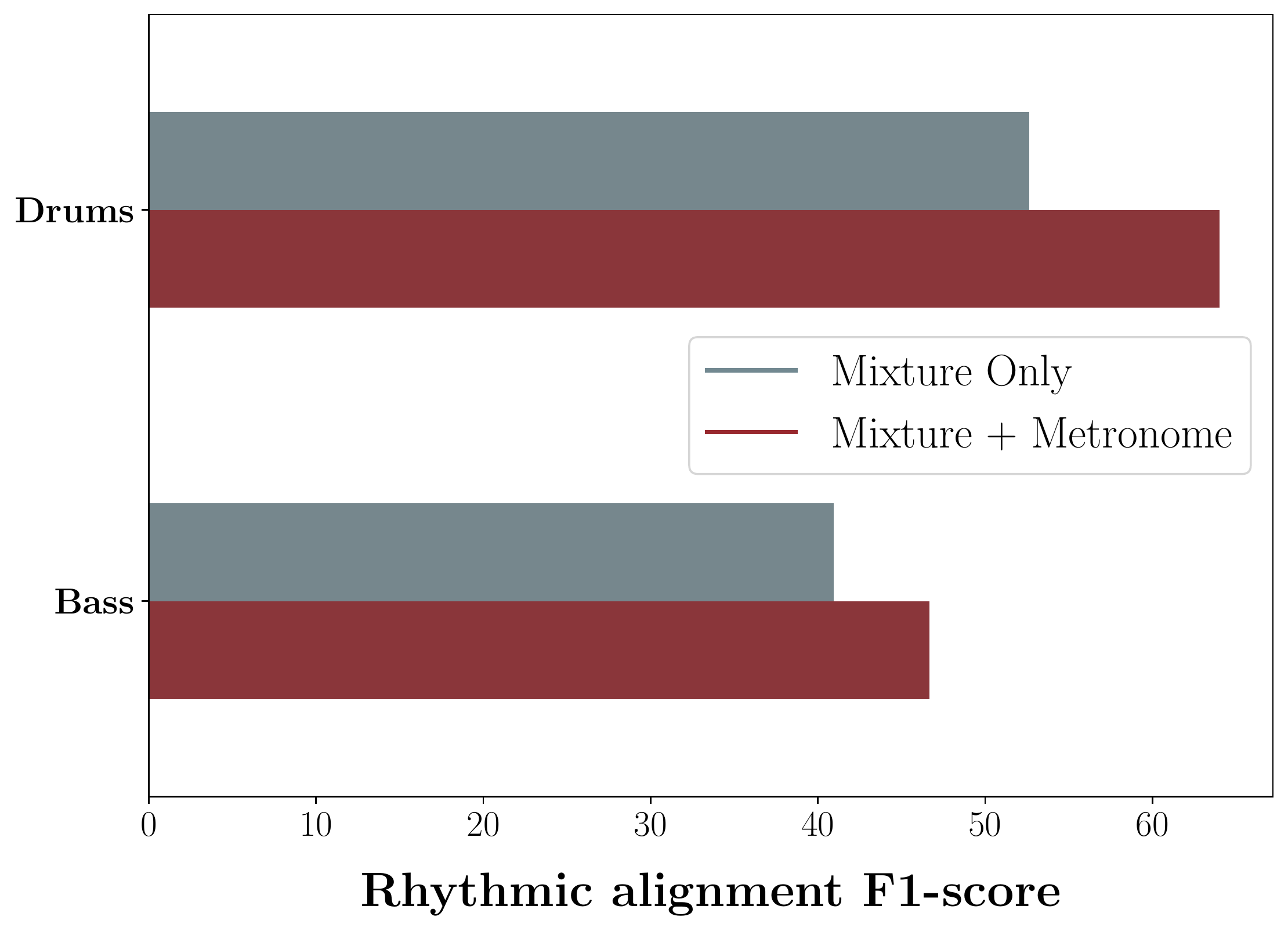}
    \caption{Comparison of rhythmic alignment when passing only a mixture as conditioning vs.\ the combination of the same mixture with a metronome track. For \stage{}\texttt{-drums}, the F1 alignment improves from $52.6$ to $64.0$, and for \stage{}\texttt{-bass} from $40.9$ to $46.8$.}
    \label{fig:rhytmic-alignment}
\end{figure}

As mentioned in \ref{subsec:overview}, we verify that, at inference time, we can condition the model on a combination of mixture and metronome, by simply summing the waveforms. Even though \stage{} has never seen such contexts during training, it is able to generalize to these conditionings and provide accompaniment generation with even greater control over the rhythmic alignment. We measure this effect on the same accompaniment generation task by comparing beat alignment (F1 score) when either conditioning on $M$ alone vs.\ on $M+B$. Results, shown in Figure \ref{fig:rhytmic-alignment} confirm that an explicit beat track helps tighten alignment while preserving coherence with the mixture.

\section{Discussion}
\label{sec:discussion}
Below we discuss a few more general takeaways from this research, and share our opinions about research on large generative transformer models. 
% When the rest of this paper becomes obsolete, probably in a few weeks, we hope for these ideas to still remain valuable for future research.

\subsection{Parameter-efficiency of single-stem fine-tuning}
As presented in Table \ref{tab:model_sizes}, \stage{} operates with roughly one-tenth of the trainable parameters of systems of comparable performance, yet it achieves similar or better results on multiple tasks. Its core strategy, training on a single stem via simple prefix-based conditioning, lets the model align closely with the given audio context without needing extra encoders or modules. Leveraging a general model like \musicgen{} and fine-tuning it on a single stem, focusing its predictive power on a simpler waveform distribution, \textbf{is a highly parameter-efficient technique}. 

These experiments suggest that stacking multiple \stage{} instances, one per instrument, can match or exceed general models' performance at a fraction of the parameters. 
This aligns with a broader trend: recent research on large language models shows that \textit{assigning subsets of parameters to specific subtasks} is often more efficient than scaling monolithic models.\begin{wraptable}[9]{r}{0.5\linewidth}
    \centering
    \resizebox{\linewidth}{!}{%
    \begin{tabular}{lc}
    \toprule
    \textbf{Model}            & \textbf{\# Params} \\ 
    \midrule
    \texttt{STAGE}       & $\sim$0.4B         \\
    \texttt{Instruct-Musicgen} & $\sim$4.7B         \\ 
    \texttt{GMSDI}        & $\sim$0.8B                  \\ 
    \texttt{SA ControlNet} & $\sim$3.8B        \\
    \bottomrule
    \end{tabular}
    }
    \caption{Considered models' trainable parameters count.}
    \label{tab:model_sizes}
\end{wraptable}%
DeepSeek V3~\cite{liu2024deepseek} exemplifies this with its use of \textit{Mixture of Experts (MoE)}, although in a different context.
We believe that, given the exponentially increasing cost, both economic and energetic, of very large models, attention to more parameter-efficient architectures should not be spared.

\subsection{Cross-attention in local vs.\ global conditioning}
Before pivoting towards prefix-based conditioning, we extensively tested cross-attention for injecting musical context into the model. While fine-tuned models captured style, mood, and harmony, they struggled with precise rhythmic alignment, even with positional embeddings designed for local dependencies. Moving the conditioning source into the input stream, processed via self-attention, resolved this issue entirely, enabling accurate alignment between conditioning and output. This behavior has rarely been documented, with \cite{liu2025fasterdiffusiontemporalattention} noting that \textit{cross-attention outputs converge to a fixed point in the first few steps}, splitting the process into semantic planning (via cross-attention), and subsequent image generation.

% \begin{quote}
%  Accordingly, the time point of convergence divides the denoising process of the diffusion models into two stages: i) an initial stage,
% during which the model relies on cross-attention to plan text-oriented visual semantics [...] and ii) a subsequent stage, during which the model learns to generate images from previous semantic planning [...].
% \end{quote}

In \cite{1360299149686305408}, there are hints of a similar intuition. We conclude that the precise behavior of cross-attention in traditional transformer decoders, and its difference in effectiveness when handling global (e.g.: the global description of an image, the meaning of a text to translate) vs.\ highly localized conditioning (e.g.: the exact positions of the beats to follow to in a musical piece) is way underexplored, and still deserves attention for future research.

\section{Conclusion}
\label{sec:conclusion}
We introduced \stage{}, a parameter-efficient single-instrument approach for accompaniment generation that extends \musicgen{} with a simple, prefix-based conditioning mechanism. By prepending an audio context to the model’s input, \stage{} effectively learns the relationship between context and accompaniment, delivering competitive or superior performance in audio fidelity, contextual coherence, and beat alignment compared to larger baseline systems, despite its compact size and minimal fine-tuning effort. We explored how enabling the model to be conditioned on either a full mixture or a simple metronome track, not only enables it to perform strictly temporally-controlled generation, but also improves its rhythmic alignment in the more general accompaniment generation task.
Future work may explore extending \stage{} to additional instruments and refining its capacity for an even more granular degree of control, further expanding its applicability in real-world music production workflows.

% \section{Acknowledgments}
% TODO

% \section{Ethics Statement}
% TODO

\bibliography{references}

\end{document}